%% file: template.tex
\documentclass[preprints,article,accept,pdftex,moreauthors]{Definitions/mdpi}

\firstpage{1}
\pubvolume{1}
\issuenum{1}
\articlenumber{0}
\pubyear{2023}
\copyrightyear{2023}
\datereceived{ }
\daterevised{ } 
\dateaccepted{ }
\datepublished{ }
\hreflink{https://doi.org/} 

\usepackage{amssymb}
\usepackage{amsmath}
\usepackage{enumitem}

\usepackage{gensymb}

\usepackage{tabularx}
\usepackage{multirow}

\Title{3DGAUnet: 3D generative adversarial networks with a 3D U-Net based generator to achieve the accurate and effective synthesis of clinical tumor image data for pancreatic cancer}

\TitleCitation{3DGAUnet: 3D generative adversarial networks with a 3D U-Net based generator to achieve the accurate and effective synthesis of clinical tumor image data for pancreatic cancer}

\Author{Yu Shi $^{1}$, Hannah Tang$^{1}$, Michael Baine$^{2}$, Michael A. Hollingsworth$^{2}$,   Huijing Du$^{1}$, Dandan Zheng$^{3}$, Chi Zhang$^{1}$, Hongfeng Yu$^{1}$}

\AuthorNames{Yu Shi, Hannah Tang, Michael Baine, Michael A. Hollingsworth, Huijing Du, Dandan Zheng, Chi Zhang, and Hongfeng Yu}

\AuthorCitation{Shi, Y.; Tang, H.; Baine, M.; Hollingsworth, M.; Du, H.; Zheng, D.; Zhang, C.; Yu, H.}

\address{%
$^{1}$ \quad University of Nebraska-Lincoln, Lincoln, NE, USA; yu.shi@huskers.unl.edu(Y.S.); htang11@unl.edu(H.T.); hdu@unl.edu(H.D.); zhang.chi@unl.edu(C.Z.); hfyu@unl.edu(H.Y.)\\
$^{2}$ \quad University of Nebraska Medical Center, Omaha, NE, USA;  mbaine@unmc.edu(M.B.); mahollin@unmc.edu(M.H.) \\
$^{3}$ \quad University of Rochester Medical Center, Rochester, NY, USA; Dandan\_Zheng@URMC.Rochester.edu(D.Z.)
}

\corres{Correspondence: zhang.chi@unl.edu(C.Z.); hfyu@unl.edu(H.Y.)}

\abstract{
Pancreatic ductal adenocarcinoma (PDAC) presents a critical global health challenge, and early detection is crucial for improving the 5-year survival rate. Recent medical imaging and computational algorithm advances offer potential solutions for early diagnosis. Deep learning, particularly in the form of convolutional neural networks (CNNs), has demonstrated success in medical image analysis tasks, including classification and segmentation. However, the limited availability of clinical data for training purposes continues to provide a significant obstacle. Data augmentation, generative adversarial networks (GANs), and cross-validation are potential techniques to address this limitation and improve model performance, but effective solutions are still rare for 3D PDAC, where contrast is especially poor owing to the high heterogeneity in both tumor and background tissues.
In this study, we developed a new GAN-based model, named 3DGAUnet, for generating realistic 3D CT images of PDAC tumors and pancreatic tissue, which can generate the inter-slice connection data that the existing 2D CT image synthesis models lack. The transition to 3D models has allowed the preservation of contextual information from adjacent slices, improving efficiency and accuracy, especially for the poor-contrast challenging case of PDAC. PDAC's challenging characteristics, such as iso-attenuating or hypodense appearance and lack of well-defined margins, make tumor shape and texture learning challenging. To overcome these challenges and improve the performance of 3D GAN models, we have developed a 3D U-Net architecture for the Generator. Thorough examination and validation across many datasets were conducted on the developed 3D GAN model to ascertain the efficacy and applicability of the model in clinical contexts. Ultimately, this approach represents a promising avenue to address the pressing need for innovative and synergistic approaches to combat PDAC. The development of this GAN-based model holds the potential for improving the accuracy and early detection of PDAC tumors, which could profoundly impact patient outcomes. Furthermore, this model has the potential to be adapted to other types of solid tumors, hence making significant contributions to the field of medical imaging in terms of image processing models.
}


\begin{document}

\input{sec_introduction}

\input{sec_methods}
\input{sec_results}

\input{sec_disbussion}
\input{sec_concolusions}

\authorcontributions{Conceptualization, C.Z. D.Z., H.D., and H.Y.; methodology, Y.S., H.T., C.Z., and H.Y.; software, Y.S., and H.T.; validation, Y.S., C.Z., D.Z., and H.Y.; data availability: M.B. and M.H.; expert contouring: M.B.; writing---original draft preparation, Y.S., C.Z., and H.Y.; writing---review and editing, Y.S., H.T., C.Z., D.Z., H.D., and H.Y.; visualization, Y.S., and H.T.. All authors have read and agreed to the published version of the manuscript.}

\funding{This research has been sponsored in part by Nebraska Collaboration Initiative grant 19-20 and 20-21,
National Institutes of Health grant 5U54GM115458-03, and the Layman Fund held at the University of Nebraska Foundation.}

\conflictsofinterest{The authors declare no conflict of interest.}

\begin{adjustwidth}{-\extralength}{0cm}

\reftitle{References}


\bibliography{references}
\PublishersNote{}
\end{adjustwidth}
\end{document}

%% file: sec_introduction.tex
\section{Introduction}

Pancreatic ductal adenocarcinoma (PDAC) causes a significant public health concern due to its delayed identification, restricted efficacy of current chemotherapeutic treatments, and poor overall prognosis. It has the most elevated fatality rate among the primary types of solid malignancies. Despite extensive clinical and research endeavors spanning decades, the one-year survival rate stands at 20\%, while the five-year survival rate remained in the single digits for a considerable time and only recently improved to 11\% \cite{ACS}. Despite the potential for a substantial increase in the 5-year relative survival rate to 42\%~\cite{SEER}, if early detection at the localized stage is achieved, there is currently a lack of definitive screening methods to identify early-stage pancreatic cancer in asymptomatic individuals reliably. 

Computed tomography (CT) is one of the primary diagnostic imaging methods. In recent years, deep-learning-based methods are increasingly perceived as versatile applications. They can directly integrate physical or semantic details into neural network architectures~\cite{wu2019survey,tandel2019review, sharif2020comprehensive, radiya2023performance}, and are employed to solve computer vision tasks in medical imaging, such as segmentation, registration, or classification of chest X-rays and tissue histopathology images \cite{baltruschat2019comparison,coudray2018classification}. For example, convolutional neural network (CNN) models have shown high feasibility in image classification tasks in both natural and medical images from 2D models to 3D models \cite{li2018deep, razzak2018deep, li2019deep}. Some similar studies were applied for pancreatic cancer classifiers to analyze and interpret features from medical imaging data~\cite{chu2021pancreatic, si2021fully}.

During the development of a deep learning model for image tasks, a substantial dataset (e.g., thousands of images) is typically needed to ensure the model converges without overfitting. Nevertheless, the availability of clinical information, particularly for PDAC, is frequently constrained by the small size of the cohorts, which presents obstacles to achieving optimal model training. Researchers have developed methods, such as data augmentation, generative adversarial networks (GAN), cross-validation, and optimization approaches like sharp-aware minimization \cite{foret2020sharpness}, to overcome the lack of training data. Generative models have demonstrated efficacy in medical image synthesis, particularly in 2D imaging modalities. Recently, researchers have developed 2D-based GAN models to generate realistic CT images of pancreatic tumors \cite{wei2022pancreatic, guan2022medical}.
Nevertheless, the utilization of 3D generative models in the context of PDAC is still constrained, and directly applying existing approaches (e.g., 3D-GAN~\cite{wu2016learning}) may not lead to desirable results for synthesizing three-dimensional CT image data specific to PDAC. PDAC tumors often exhibit subtle imaging features because they can be iso-attenuating or hypodense compared to the surrounding pancreatic tissue, making them difficult to distinguish visually. Additionally, PDAC tumors may lack well-defined margins, making differentiating them from normal pancreatic parenchyma challenging. Therefore, developing efficient techniques for enhancing 3D PDAC tumor datasets is crucial to facilitate the progress of deep learning models in addressing PDAC.

In this work, we develop a GAN-based tool capable of generating realistic 3D CT images depicting PDAC tumors and pancreas tissue. To overcome these challenges and make this 3D GAN model have a better performance, we have developed a 3D U-Net architecture for the Generator.  The application of 3D U-Net in medical picture auto segmentation has shown appropriate and superior results. Notably, this is the first instance of its integration into GAN models. This 3D GAN model generates volumetric data of PDAC tumor tissue CT images and healthy pancreas tissue CT images separately, and a blending method was employed to create realistic final images. Thorough examination and validation across many datasets have been conducted on the developed 3D GAN model to ascertain the efficacy and applicability of the model in clinical contexts. We evaluate the effectiveness of our approach by training a 3D CNN model with synthetic image data to predict 3D tumor patches. A software package, 3DGAUnet, was developed to implement this 3D generative adversarial network with a 3D U-Net based generator for tumor CT image synthesis. This package has the potential to be adapted to other types of solid tumors, hence making significant contributions to the field of medical imaging in terms of image processing models. This software package is available at https://github.com/yshi20/3DGAUnet.





%% file: sec_methods.tex
\section{Materials and Methods}

\begin{figure}[t]
\begin{center}
\includegraphics[width=1.0\linewidth]{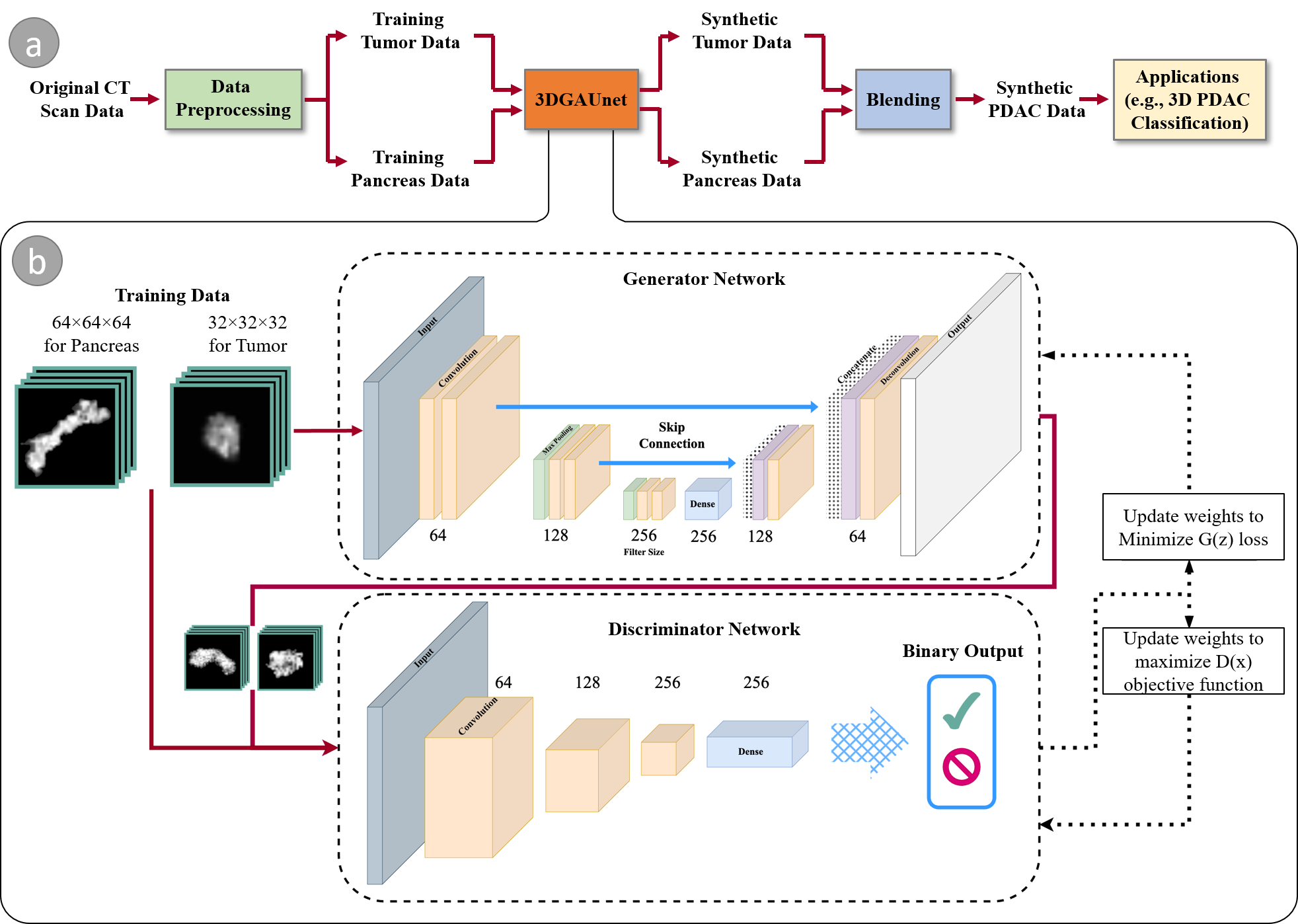}
\end{center}
\caption{ \textbf{An overview of our method.}  (a) the workflow components and (b) the architecture of our GAN-based model, 3DGAUnet, consisting of a 3D U-Net-based generator network and a 3D CNN-based discriminator network to generate synthetic data.}
\label{fig:workflow}
\end{figure}

Figure~\ref{fig:workflow} (a) illustrates the overall workflow of our proposed method. Given a set of PDAC CT images that can be acquired through different sources, we first design and conduct data preprocessing on these raw image data to tackle data heterogeneity and generate normalized and resampled volume data for tumor tissues and pancreas. These preprocessed datasets are then used as the training set and fed into 3DGAUnet, a new 3D GAN model developed in this work for tumor CT image synthesis. After the tumor and pancreas types are learned independently via 3DGAUnet, the corresponding synthetic data can be generated. 

To effectively combine these synthetic tissues, we evaluate three blending methods and identify the most suitable technique for PDAC tumor CT images. Given that the pancreas is a parenchymal organ, the relative location of the tumor tissue is found to be less significant. As a result, the focus is primarily on blending the different tissue types seamlessly and realistically to ensure accurate and reliable results for diagnosing PDAC tumors in CT images.  

We have evaluated the usability of the synthetic data by applying it to the diagnosis task. For this purpose, we employ a 3D CNN classifier capable of taking 3D volumes as input, which is an improvement over traditional classification tools that only use individual slices and overlook the inter-slice information.

By integrating the synthetic data, we addressed common challenges encountered in real-world scenarios, such as the small size of the data set and imbalanced data. The addition of synthetic samples helped to improve the model's performance and mitigate issues related to imbalanced datasets.

\subsection{PDAC Tumor 3D CT Image Data Preprocessing}
\label{sec:preprocessing}

We use a training data set for the GAN model consisting of PDAC CT images of 174 patients from two sources. One is from the University of Nebraska Medical Center (UNMC) rapid autopsy program (RAP). This data set has 71 de-identified patient data with the tumor contour labeled by clinical professionals (UNMC IRB PROTOCOL \#127-18-EP). The other is the Medical Segmentation Decathlon pancreases training data \cite{antonelli2022medical}, which has 103 volumetric images with the segmentation mask of tumors and pancreas.

Normalization and resampling are essential for these raw image data, as the images were obtained from various instruments with distinct configurations. By resampling all volumetric data to 1mm isotropic voxel spacing, each pixel on the image will represent the same physical distance along each axis. For normalization, the window level threshold is an important measurement. In CT imagery, the Hounsfield unit (HU) is used as a dimensionless unit to measure the radio density and quantify the tissues within the body. It is calculated based on a linear transformation of X-ray's baseline linear attenuation coefficient, where distilled water is defined to be zero HU and air is defined as -1000 HU~\cite{hounsfield1980computed}. Similar HU values across different studies indicate the same type of tissue. However, calculating HU values for grayscale images with different window-level settings can lead to different visual appearances. In this study, the original images have HU values ranging from -408 to 1298. For normalization, we map HU values to the range of -100 to 170 for abdominal soft tissues based on the advice of an experienced radiologist.

The original image data from UNMC have metal markers that cause extremely high HU values and deflect the X-ray beam, causing the sounding tissue to have a higher HU value. To counter this defect, pixels above 200 HU were replaced with the mean HU value of the entire pancreas captured in each CT scan. Since the pancreas has an irregular shape, the tumor region and surrounding pancreas tissues that fill a cube with \(64\times64\times64\) pixels were kept as the field of interest. This will make the GAN model learn the texture instead of learning the premier of the pancreas organ that varies among the different patients. 
After the data preprocessing step, the training volumetric data are \(32\times32\times32\) for tumor tissue and \(64\times64\times64\) pancreas cubes in grayscale.


\subsection{3DGAUnet: 3D U-Net Based GAN Model}


We devised a 3D U-Net based GAN model, named 3DGAUnet, to synthesize 3D pancreas tumor and tissue images. Figure~\ref{fig:workflow} (b) shows our model architecture.
At a high level, our model follows a typical GAN model that comprises two primary components, a generator $G$ and a discriminator $D$. The generator creates synthetic samples, while the discriminator differentiates between synthetic and natural samples. They compete in an adversarial game to improve the generator's ability to generate genuine samples and the discriminator's ability to identify them. The aim is to produce synthetic samples that closely approximate their natural counterparts. This process can be defined as a min-max optimization task
\begin{equation}
min_G max_D L_{GAN}, 
\end{equation}
and $L_{GAN}$ is defined as:
\begin{equation}
L_{GAN} = \mathbb{E}_{x\sim p_{data}} [\log D(x)] + \mathbb{E}_{z\sim p_z} [\log (1 - D(G(z)))]
\end{equation}
where $\mathbb{E}$ is the cross-entropy of the binary classifier of the discriminator.
The task of the generator $G$ is to minimize the generator loss to generate synthesized images that cannot be distinguished by the discriminator $D$:
\begin{equation}
min_G L_{GAN} = min_G \mathbb{E}_{Z\sim p_Z} [\log (1 - D(G(z)))]
\end{equation}
The task of the discriminator $D$ is to separate real images and synthesized images better:
\begin{equation}
max_D L_{GAN} =  max_D \,\{\mathbb{E}_{x\sim p_{data}} [\log D(x)] + \mathbb{E}_{z\sim p_z} [\log (1 - D(G(z)))]\}. 
\end{equation}

In our 3D GAN model, we employ a 3D U-Net-based structure for a 3D GAN generator and a 3D CNN-based classifier as the discriminator. The 3D U-Net structure has proven advantages for effectively capturing both global and local structures, such as for CT image auto-segmentation for tissues like the pancreas~\cite{oktay2018attention}. We have developed a 3D U-Net structure for 3D image synthesis, and it is the first time that a 3D U-net structure is used for a Generator in a GAN model. Each convolution layer in this model has a kernel size of \(3\times3\times3\), a stride of 2, and a ReLU activation. The skip connections allow the low-level information to be passed to the up-sampling stacks to avoid the gradient vanishing problem.

The discriminator is to identify if the input image is synthesized images from the generator's output. It has three 3D convolutional blocks, and each block starts with a 3D convolution with a kernel size of \(2\times2\times2\) and sides of 1, followed by a 3D Maxpooling layer with a pool size of 2, and batch normalization. After three 3D convolutional blocks, it flattened to a fully connected dense layer, and binary output was generated by the sigmoid function.

Our approach can be used to train 3D models of tumors and pancreatic healthy tissue separately. The training procedure is optimized with respect to discriminator loss. A total of 500 3D tumor and pancreas volumes are synthesized. By inserting tumor tissue into the pancreas cube, a 3D volume of the tumor with surrounding pancreas tissues can be generated.

\subsection{Blending to Create PDAC Tissues}

Because the tumor and pancreas tissue volumes have been generated separately, it is essential to consider how to combine the tumor and pancreas volumes to create realistic tumorous pancreas tissue. We have experimented with and compared three blending methods for merging tumor and tissue volumes of a pancreas. The first method (namely Blend I) is a straightforward copy-and-paste operation that uses the tumor voxels to replace the corresponding pancreas tissue voxels. Our second and third methods (namely Blend II and Blend III, respectively) are inspired by DeepImageBlending, a deep learning technique that improves Poisson image blending \cite{zhang2020deep}. Deep image blending is a two-stage image blending algorithm. First, it generates a seamless boundary for the source region to eliminate visible seams. Then, it refines the source region by matching styles and textures with the target image. The algorithm uses a differentiable loss function based on the Poisson equation and can handle various image styles, including stylized paintings. It achieves visually consistent blending without relying on training data. Our Blend II and Blend III methods are the first and second stages of the tool, respectively. The idea of comparing the two stages is because, in a natural image blending task, the object should look natural and share a similar style with the background image, but this might not be true for a CT image. Unlike natural images that are acquired from the light reflection of the objects, CT images are created by recording the X-ray beam attenuation from different directions, and therefore, a presumption of a similar style may not be valid. To find the best blending method, we have compared the three blending methods by visual inspection and Fréchet Inception Distance (FID) values. After synthetic tumors and healthy pancreas tissues have been directly output by our 3D GAN model, they are blended to generate synthetic PDAC tissues with the best blending method. The comparison of the blending methods will be provided in Section~\ref{sec:blending_compare}.


\subsection{Evaluation of Synthesized Images}

The performance of our developed 3DGAUnet model is evaluated both qualitatively and quantitatively. We visualize the generated volumes with 2D cross slices and 3D volume rendering for qualitative evaluation~\cite{kikinis20133d}. For quantitative evaluation, we use FID values on 2D slices \cite{heusel2017gans}.

We propose a 3D evaluation metric, called Fréchet 3D Distance (F3D), for comparing the Fréchet distance of the activation layer from a 3D CNN network to compare the quality of the 3D GAN model. The distance $d$ is calculated as:
\begin{equation}
d^2 = ||\mu_1 - \mu_2||^2 + Tr(C_1 + C_2 - 2\sqrt{C_1 C_2}),
\end{equation}
where the $\mu_1$ and $\mu_2$ are the feature-wise means of the real and synthesized images, $C_1$ and $C_2$ are the covariance matrix of their feature vector for the real and synthesized images, and $Tr$ is the trace linear algebra operation that is the sum of the elements along the main diagonal of the square matrix.

To calculate the $\mu$ and $C$, we need the feature vector from the last pooling layer out of a pre-trained neural network. The original FID uses a trained Inception V3 model \cite{szegedy2016rethinking}. Our approach, instead, uses a 3D CNN with 17 layers, including four 3D convolutional blocks with a fixed random state 42, and the feature vector is the flattened layer after the last convolutional block, having a length of 512. Samples are compared in batches, and then the $\mu$ and $C$ can be calculated with the matrix that consists of the feature vector from each sample.

In addition, the quality of generated images by our developed 3DGAUnet model was evaluated with the squared Maximum Mean Discrepancy (MMD$^2$) that employs kernel functions in the reproducing kernel Hilbert space to quantify the discrepancy between two distributions~\cite{gretton2012kernel}. In this study, we also use pair-wise Multi-Scale Structural Similarity (MS-SSIM) to assess the diversity of images generated by our 3DGAUnet model. MS-SSIM is a metric that quantifies the perceptual diversity of the generated images by calculating the mean of MS-SSIM scores for pairs of these images~\cite{odena2017conditional}. This measurement allows us to evaluate the level of variation and dissimilarity among the generated samples, providing insights into the model's ability to produce diverse and distinct images.

\subsection{3D CNN PDAC Classifier}

One of our objectives to create synthetic data is to improve the performance of PDAC tumor identification. Currently, limited available data is an obstacle. To test our developed 3D GAN model, we build and train a 3D CNN classifier using the synthetic data generated by our 3DGAUnet model. 

We develop a 17-layer 3D CNN model~\cite{tran2015learning} to test if a 3D volumetric input is healthy pancreas tissue or has a tumor. 
The 3D CNN classifier has four 3D convoluted blocks (Conv 3D), with the first block consisting of 64 filters followed by 128, 256, and 512 filters, all with a kernel size of \(3\times3\times3\). Each Conv3D layer is followed by a max-pooling (MaxPool) layer with a stride of 2, ReLU activation, and batch normalization layer (Batch Norm). This 3D CNN model has four Con3D-MaxPool-BatchNorm blocks and intends to capture the visual features from coarse to fine.
The final output first flattens the output of the last convolutional block and passes it to a fully dense layer with 512 neurons. A dropout layer with a tunable dropout rate is followed to prevent overfitting. The output is then passed to a 2-neuron dense layer with a sigmoid function for binary classification output. Because the input dimension is \(64\times64\times64\), as a relatively simple task, the architecture of the classifier is designed in a simple way to avoid the overparameterization problem with 1,351,873 learnable parameters.

The binary classification performance is calculated from the confusion matrix. Given that $TP$, $TN$, $FP$, and $FN$ correspond to true positive, true negative, false positive, and false negative, respectively, results are measured by precision $TP / (TP + FP)$, recall $TP / (TP + FN)$,  true positive rates $TPR = TP / (TP + FN)$, and the false positive rates $FPR= FP / (FP + TN)$.  The area under the curve (AUC) is calculated from the Receiver Operating Characteristic (ROC) curve, which is plotted as true positive rates against the false positive rates under different cutoffs or as the precision against recall.

%% file: sec_results.tex
\section{Results}



\begin{figure}[t]
\centering
$\begin{array}{cc}
\includegraphics[width=0.49\linewidth]
{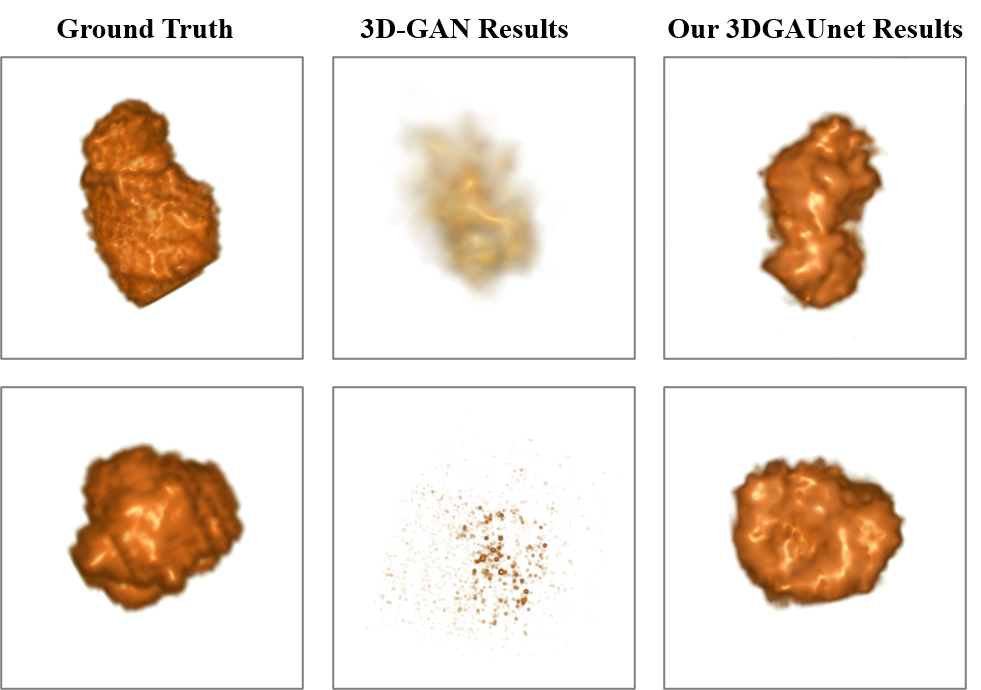} &
\includegraphics[width=0.49\linewidth]
{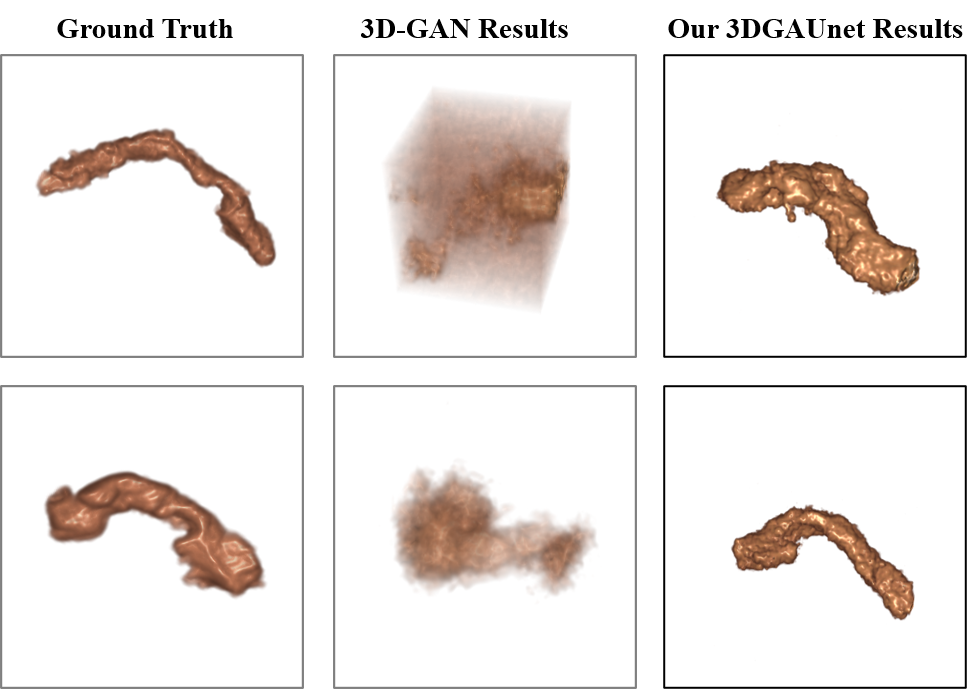}
\\
\mbox{{\small (a) Turmor Examples}} & \mbox{{\small (b) Pancreas Examples}}
\end{array}
$
\caption{\textbf{Examples of  3D volumes obtained from different methods}. Examples of 3D volume data of tumor (a) and pancreas (b) from different methods. In each set of examples, the left, middle, and right columns correspond to ground truth data, synthetic data generated by 3D-GAN, and synthetic data generated by our 3DGAUnet, respectively. All 3D volumes are shown in volume rendering.}
\label{fig:tumor_pancreas_coompare}
\end{figure}

\subsection{3D Volumetric Tissue Data Generation}

We train our 3DGAUnet model separately using PDAC tumor and healthy pancreas data. They are referred to as the tumor and pancreas models. The tumor model is trained using PDAC tumor data, including 174 volumetric tumor data in Nifty format. The pancreas model is trained using healthy tissue data, including 200 volumetric data in Nifty format. Both input datasets result from the preprocessing steps outlined in Section~\ref{sec:preprocessing}.

Image augmentation, including image flipping and rotation, has been performed on the training data. The augmented data for each volume is generated by rotating each volume on three axes in 12\degree, 24\degree, 36\degree, 48\degree, and 72\degree increments. All images for the tumor model have been resampled to 1 mm isotropic resolution and trimmed to \(32\times32\times32\) size.
All images for the pancreas model have been resampled to 1 mm isotropic resolution and trimmed to \(64\times64\times64\) size with pancreas tissue filling the entire cube.

The training procedure of any GAN model is inherently unstable because of the dynamic of optimizing two competing losses. For each model in this study, the training process saved the model weights every 20 epochs, and the entire model is trained for 2000 epochs. The best training duration before the model collapses is decided by inspecting the generator loss curve and finding the epoch before the loss drastically increases. We have trained our models with one NVIDIA RTX 3090 GPU. 
The optimal parameter set is searched within a parameter space consisting of batch size and learning rate, where the possible batch sizes include 4, 8, 16, and 32, and the possible learning rates include 0.1, 0.01, 0.001, 0.0001, and 0.00001.

\begin{figure}[t]
\centering
$\begin{array}{cc}
\includegraphics[width=0.49\linewidth]
{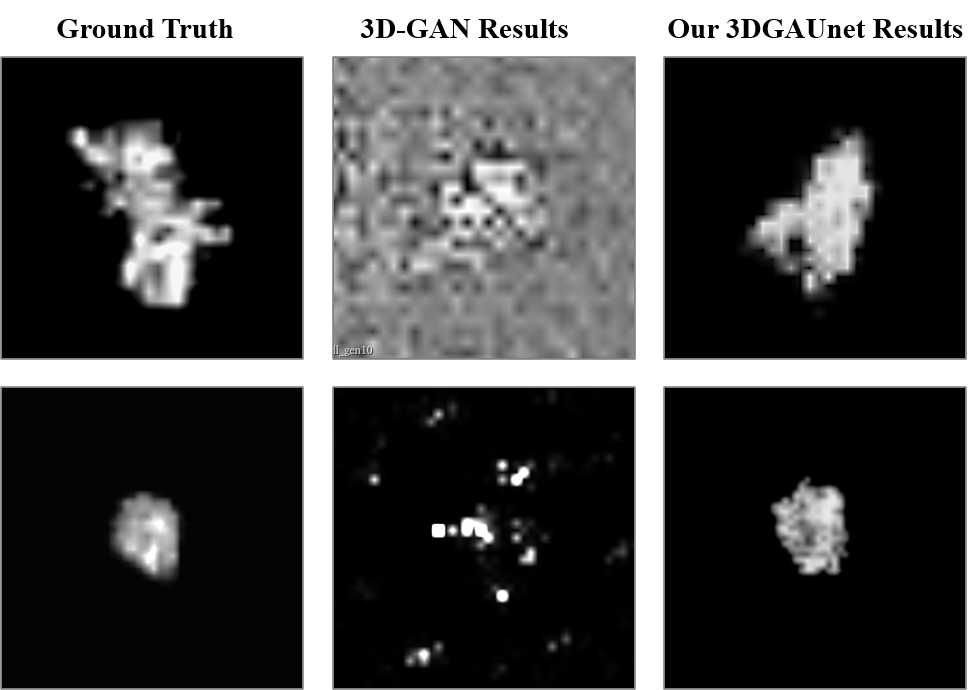} &
\includegraphics[width=0.49\linewidth]
{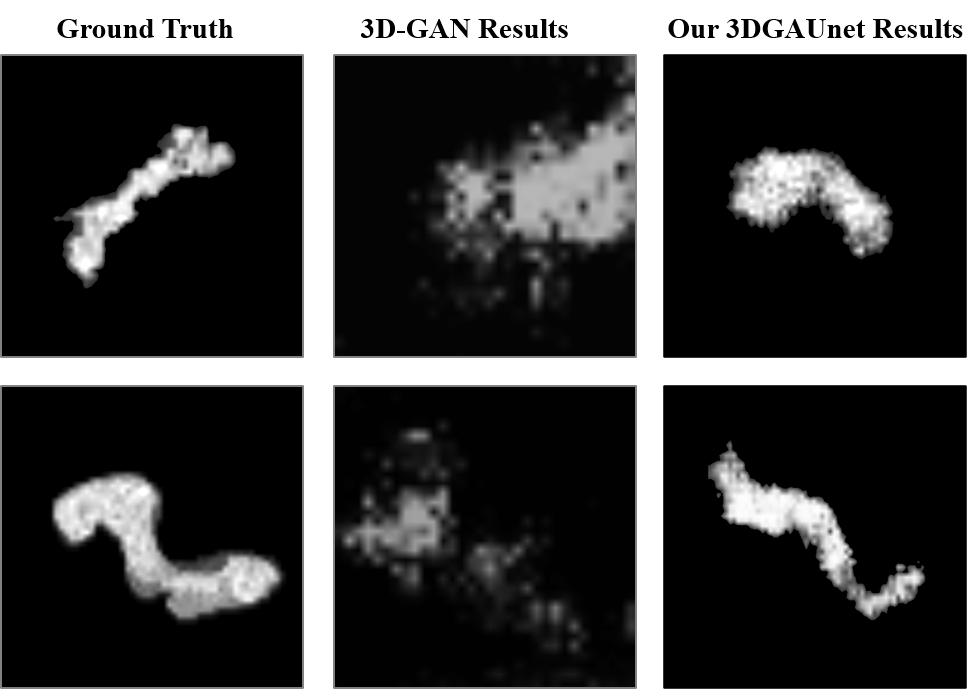}
\\
\mbox{{\small (a) Turmor Examples}} & \mbox{{\small (b) Pancreas Examples}}
\end{array}
$
\caption{\textbf{Examples of 2D slices in 3D volumes obtained from different methods.} Examples of 2D in 3D volumes of tumor (a) and pancreas (b) from different resources. In each set of examples, the left, middle, and right columns correspond to ground truth data, synthetic data generated by 3D-GAN, and synthetic data generated by our 3DGAUnet, respectively.}
\label{fig:2D-compare}
\end{figure}

A set of 500 synthetic volumetric data is generated by the tumor and pancreas models separately. We have first conducted a qualitative comparison between the training image sets and the synthetic image sets. We use volume rendering to visualize these datasets to inspect 3D results. Figure~\ref{fig:tumor_pancreas_coompare}(a) shows examples of ground truth tumor volumes, synthetic tumors generated by the existing technique 3D-GAN~\cite{wu2016learning}, and synthetic tumors generated by our 3DGAUnet. We can see that once we train our 3DGAUnet based on the group truth inputs, our model can generate synthetic tumors carrying realistic anatomical structure and texture and capturing overall shape and details. Nonetheless, 3D-GAN either produces unsuccessful data or fails to generate meaningful results to capture the tumor's geometry. 
Figure~\ref{fig:tumor_pancreas_coompare}(b) shows examples of group truth pancreas volumes and synthetic pancreas volumes generated by 3D-GAN and our 3DGAUnet. By comparing the generated pancreas volumes with real medical images, we can see that our 3DGUnet can effectively synthesize 3D pancreas to resemble actual anatomical structures. However, it is hard for 3D-GAN to generate anatomically plausible results, and certain ambient noise has been perceived in generated volumes. 
We further examine the interior structures of volumes generated by our 3DGUnet. Figure~\ref{fig:2D-compare} shows the 2D slices of ground truth, 3D-GAN, and our 3DGAUnet images from both tumor and pancreas models.
The synthetic data produced by our 3DGAUnet model exhibits a high degree of fidelity to the ground truth in terms of both internal anatomical structure and texture compared to 3D-GAN. In certain instances, 3D-GAN fails to produce meaningful outcomes.


However, we can also observe marginal defects among pancreas generation that tiny tissue surrounding the main tissue generated in the center. The defects are likely due to the irregular shape, different sizes, and direction of the pancreas, then the gradient learned from the input images batches at certain locations turning into the noises.



In addition, we conduct a quantitative assessment of the outcomes produced by 3D-GAN and our 3DGAUnet models. A sample of 100 ground truth volumes and 100 synthetic volumes was selected randomly. In order to assess the quality of synthetic volumes, certain 2D image metrics, such as slice-wise FID and slice-wise PSNR, are computed. Besides the 2D metrics, 3D metrics, such as batch-wise F3D, MMD$^2$, and SSIM, are calculated on the randomly selected volumetric data. Table~\ref{table:2D-metrics} shows the values of 2D image metrics, slice-wise FID and slice-wise PSNR, on the Sagittal (Sag), Axial (Ax), and Coronal (Cor) planes to estimate the quality of synthetic volumes, where scores are calculated using the center slice from 100 synthesized volumes and 100 ground truth volumes from the tumor model and pancreas model separately. Table~\ref{table:3D-metrics} shows the values of 3D volume metrics, batch-wise F3D, the MMD$^2$, and the MS-SSIM. From the results, we observe that our 3DGAUNet outperforms 3D-GAN in all metrics, suggesting that 3DGAUNet excels at capturing the 3D shape and texture characteristics for both tumor and pancreas compared to 3D-GAN. In addition, it is evident that all the quantitative metrics in the pancreas model are better than the tumor model, especially on FID and F3D. This is probably caused by the difference in the training tasks, where the tumor model needs to learn both the texture and shape of the tumor, but the pancreas model is trained with pancreas-filled cubes to learn the texture of the pancreas image mostly.

\begin{table}[h]
\caption{Performance based on 2D Image Quality Metrics.}
\newcolumntype{Y}[1]{>{\centering\arraybackslash}X>{\hsize=#1\hsize\arraybackslash}X}
\begin{tabularx}{\linewidth}{>{\centering\hsize=0.9\hsize\arraybackslash}X>{\centering\hsize=1.1\hsize\arraybackslash}X>{\centering\hsize=0.9\hsize\arraybackslash}X>{\centering\hsize=0.8\hsize\arraybackslash}X>{\centering\hsize=0.9\hsize\arraybackslash}X>{\centering\hsize=1.1\hsize\arraybackslash}X>{\centering\hsize=1.1\hsize\arraybackslash}X>{\centering\hsize=1.2\hsize\arraybackslash}X}
\hline
Tissue & Model & FID-Sag & FID-Ax & FID-Cor & PSNR-Sag & PSNR-Ax & PSNR-Cor \\ \hline
\multirow{2}{*}{Tumor} & 3D-GAN & 249.32 & 262.18 & 244.27 & 20.10 & 18.63 & 19.49 \\
 & 3DGAUNet &  198.23 & 202.44 & 188.66 & 16.52 & 17.76 & 17.16 \\
\hline 
\multirow{2}{*}{Pancrease} & 3D-GAN & 293.62 & 342.60 & 335.20 & 18.20 & 16.31 & 14.05\\
 & 3DGAUNet &  287.75 & 435.72 & 327.41 & 12.73 & 7.21 & 9.42 \\ \hline
\end{tabularx}
\label{table:2D-metrics}
\end{table}

\begin{table}[h!]
\caption{Performance based on 3D Image Quality Metrics.}
\newcolumntype{C}{>{\centering\arraybackslash}X}
\begin{tabularx}{\linewidth}{CCCCC}
\hline
Tissue & Model & F3D & MMD$^2$ & MS-SSIM \\ \hline
\multirow{2}{*}{Tumor} & 3DGAN & 472.64 &  5571.90 & 0.86 \\
 & 3DGAUNet & 271.31 & 5327.32 & 0.81 \\
\hline  
\multirow{2}{*}{Pancrease} & 3DGAN & 889.40 & 8924.39 & 0.83 \\
          &3DGAUNet & 872.33 & 9122.40 & 0.77 \\ \hline
\end{tabularx}
\label{table:3D-metrics}
\end{table}

\subsection{3D Volumetric Data Blending}
\label{sec:blending_compare}

From the synthesized data, we select 100 pairs of synthetic tumors and synthetic pancreas tissue volumes. The paired data are used to evaluate three previously introduced blending methods. Quantitative evaluation is conducted by comparing the FID Score, while qualitative evaluation is conducted by visualizing the 2D slices. 100 random abdominal CT images are cropped into 64 $\times$ 64 $\times$ 64 cubes as a negative sample set, and are used to compare 3D metrics with the blended volumes.
 
\begin{figure}[t]
\centering
\includegraphics[width=1.0\linewidth]{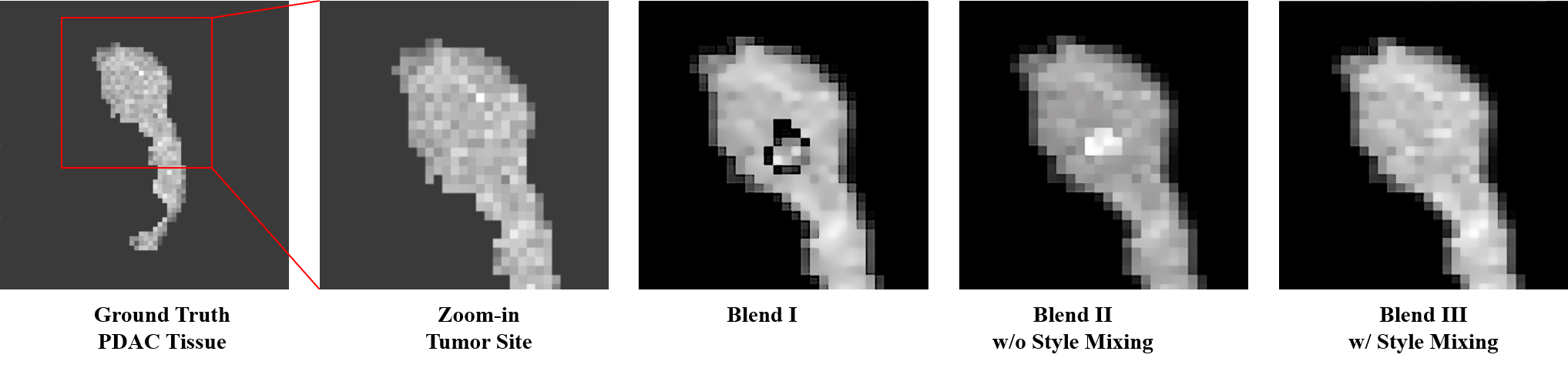}
\caption{\textbf{The comparison of blending methods.} The two left images show the details of the tumor site texture in the ground truth data as a visual reference. The rest of the images show the blend of a tumor into the healthy pancreas tissue with different blending methods. We can observe that Blend III has the best visual similarity.}
\label{fig:blending}
\end{figure}

Figure~\ref{fig:blending} shows the 2D slices of the blended images with three different blending methods. It is evident that the direct copy-and-paste approach consistently yields the least favorable results. The reason is that the tumor object is extracted with a simple threshold of pixel value, and therefore, the boundary of the tumor tissue may not be as precise as needed. One can spot black pixels randomly appearing around the boundary of the tumor tissue and pancreas tissue. Whereas Blend III is visually closer to the ground truth tumor site and, on average, has a lower slice-wise FID score. Therefore, we employ Blend III as the blending method for the developed 3D GAN model. Table~\ref{table:blending-compare} compares the slice-wise FID values among the three blending methods and clearly shows that Blend III has achieved the best slice-wise FID values.

\begin{table}[h]
\caption{Slice-wise FID Values of Blending Methods.}
\newcolumntype{C}{>{\centering\arraybackslash}X}
\begin{tabularx}{\linewidth}{CCCC}
\hline
Blending Methods & FID-Sag & FID-Ax & FID-Cor \\ \hline
Blend I & 42.10 & 40.26 & 32.94 \\
Blend II & 21.01 & 35.82 & 12.62 \\
Blend III & 13.21 & 13.88 & 10.06 \\ \hline
\end{tabularx}
\label{table:blending-compare}
\end{table}



\subsection{Enhanced Training Dataset with Synthesized Data to Improve 3D PDAC Tumor Classification}

We train a 3D CNN classifier using two different dataset configurations, with or without adding the synthetic data and compare the performance of the binary classification of PDAC between them. Adding synthesized data to the training data can enlarge the training data and reduce the imbalance between positive and negative categories because, in practice, it is usually more difficult to access PDAC patient images than healthy pancreas images.   We have a total of 174 tumorous pancreas images and 254 healthy pancreas images, all from real-world CT scans. All the input images are 1mm isotropic resolution CT volume and trimmed to the size of \(64\times64\times64\). Out of all data, 35 tumorous pancreas images and 51 healthy pancreas images, i.e., 20\% of all data, are saved as the test dataset. We have two configurations for the training dataset. The first configuration (namely Config I) only contains real data for training, i.e., 139 tumorous pancreas images and 203 healthy pancreas images. The second configuration (namely Config II) included both the training set from Config I and synthesized data, i.e., 114 synthetic PDAC images and 50 synthetic healthy pancreas images. Both training datasets have the same test dataset for comparison. Config I is a baseline for real-world imbalanced data having a smaller size, and Config II uses synthetic data to balance the entire dataset. Table~\ref{table:datasets} summarizes the training used in two different configurations.  
For data augmentation, each CT scan underwent a random rotation along a single axis. The rotation angle was randomly selected from a set of options: 5\degree, 10\degree, 20\degree, and 40\degree. The direction of rotation, either clockwise or counterclockwise, was also randomly determined.
The best parameters of each model are found with a grid search of a parameter space consisting of batch size and learning rate, where the batch sizes include 8, 12, and 16, and the learning rates include 0.001, 0.0001, and 0.00001.  All the models are trained with one NVIDIA RTX 3090 GPU and are validated by a 3-fold cross-validation.



\begin{table}[t]
\caption{Dataset Configurations for Classifier Experiments.}
\begin{tabularx}{\linewidth}{>{\hsize=0.4\hsize}X>{\hsize=1.6\hsize}X}
\hline
 & Training Set\\ 
 \hline
 \multirow{2}{*}{Config I} & 139 True PDAC \\
 & 203 True Healthy Pancreas \\
 \hline
  \multirow{2}{*}{Config II} & 139 True + 114 synthesized PDAC \\
 & 203 True + 50 synthesized Healthy Pancreas \\
 \hline
\end{tabularx}
\label{table:datasets}
\end{table}

\begin{figure}[t]
\centering
$\begin{array}{cc}
\includegraphics[width=0.5\linewidth]{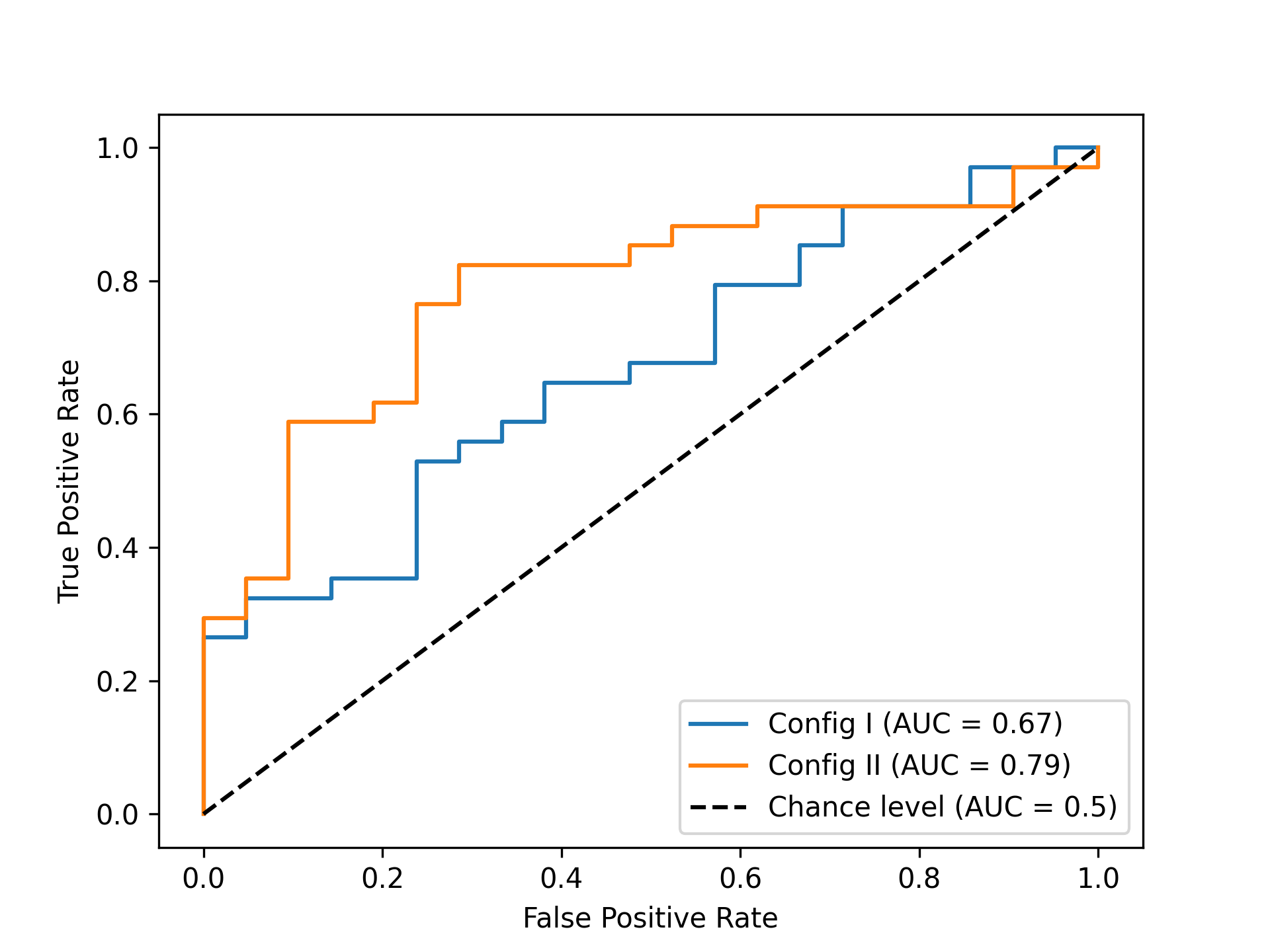}  &
\includegraphics[width=0.5\linewidth]{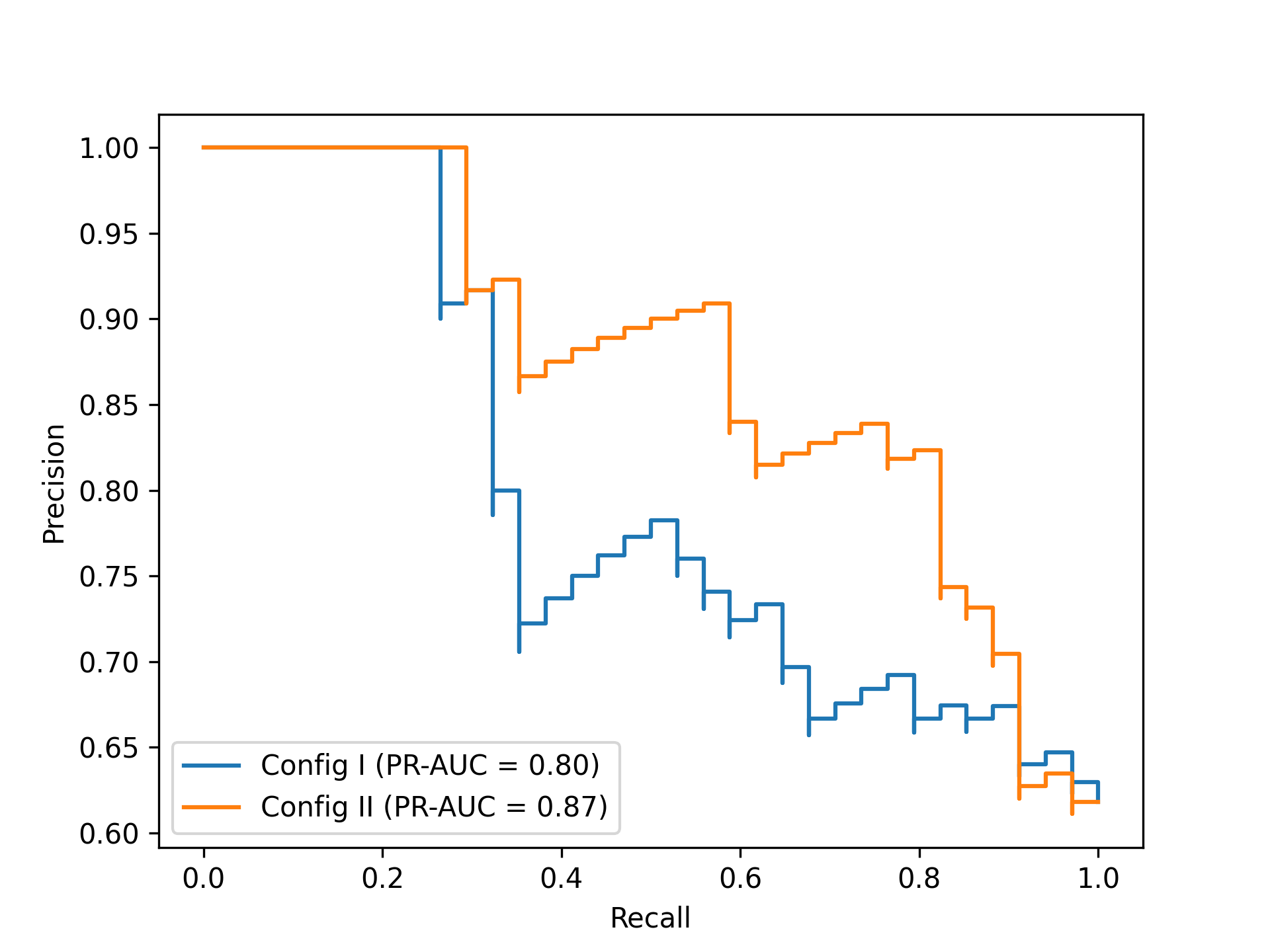} \\
\mbox{\small(a)}     & \mbox{\small(b)} 
\end{array}
$
\caption{ \textbf{3D CNN classifier performance}. (a) ROC curves and (b) PR curves with two configurations of training datasets. }
\label{fig:classification-roc}
\end{figure}

Figure~\ref{fig:classification-roc} shows the receiver operating characteristic (ROC) and precision-recall (PR) curves for the classification models. Config I has an ROC AUC (Area Under the Curve) value of 0.67 and a PR AUC value of 0.80, while Config II has an ROC AUC value of 0.79 and a PR AUC value of 0.87. The analysis of the results indicates that, as the training dataset is enlarged and the training data in the two classes become more balanced, there is an observable increase in the AUC and Precision-Recall metrics. This finding implies that including synthesized data to solve issues related to training data quantity and class imbalance has a beneficial effect on the performance of the PDAC classifier. It is worth acknowledging that, despite advancements in utilizing large amounts and balanced training data, there is considerable room for enhancing the classifier's overall performance. This might be achieved using a purpose-built 3D CNN model or by further refining the training methodology. By leveraging synthesized data, conducting extra research and analysis on the identification of supplementary components can improve the performance of classifiers and yield superior outcomes.

%% file: sec_disbussion.tex
\section{Discussion}

In this work, we develop a 3D GAN model, 3DGAUnet, for tumor CT image synthesis, compare different blending methods for CT image synthesis, and explore the impact of our synthesis method on a real-world 3D CNN classifier for tumor diagnosis.
3DGAUnet is specifically designed for synthesizing clinical CT images by combining the 3D U-Net architecture with GAN principles to generate realistic 3D CT scans of clinical data. To ensure accuracy, we train the model using 3D image data of both tumor tissue and healthy pancreas tissue. 
The quality of the synthesized images has been rigorously evaluated using both qualitative and quantitative methods. The generated images have demonstrated a more realistic texture than general 3D-GAN with the CNN-based generator and exhibited the advantage of preserving spatial coherence better than 2D methods.
One notable feature of our 3DGAUnet model is its ability to learn the inter-slice gradient, contributing to the overall realism of the generated data. The model has also showcased consistent 2D FID values across all three axes, further affirming its capability to produce high-quality 3D images.
%
 
3DGAUnet uses preprocessed, fixed-size image cubes. Preprocessing still requires a significant amount of human labor and judgment, such as eliminating defects caused by high-density material markers and creating standardized volumes for each training dataset. All training datasets must also be manually annotated by medical professionals. More automatic methods would be desirable to reduce the largest cost of acquiring data for model training.

The blending method used in this work can insert the tumor image into the background tissue image at a fixed location. For the use case involving mesenchymal organs, the various locations of the tumor have distinct anatomical meanings, and the model must also acquire this information. One possible extension is to include a segmentation module that can extract the features of each tissue type or organ from the original CT scans. By adopting this approach, the necessity of taking the blending technique into account might be eliminated, hence potentially mitigating the occurrence of faults.

The F3D score, which we implement in this work for evaluation, is a naive extension of the original FID metric, and the activation vectors have been extracted from an untrained, cold-started 3D CNN model. The stability of the F3D score in different image domains remains untested. In the future, it is necessary to have an implementation of a benchmark dataset and rigorous testing procedures to establish a standardized measure.

The 3D CNN classifier simplifies the clinical diagnosis issue into a binary classification challenge because only healthy normal pancreas images and pancreas images contain a tumor contrast. Multiple conditions may co-exist with the pancreas, such as non-cancerous lesions, inflammatory conditions or metastatic lesions, and vascular abnormalities. These obstacles remain untouched and will increase the cost of building such a model drastically due to a lack of high-quality, well-annotated data.

%% file: sec_concolusions.tex
\section{Conclusions}

3DGAUnet model represents a significant advancement in synthesizing clinical tumor CT images, providing realistic and spatially coherent 3D data, which holds great potential for improving medical image analysis and diagnosis. In the future, we will continue to address problems associated with the topic of computer vision and cancer. We would like to investigate reasonable usages of synthetic data and evaluations of data quality and usability in practice, for example, their effectiveness in training reliable classification models.